\newif\ifdoubleblind
\newif\ifacm
\acmtrue

\ifacm
	\documentclass[sigconf]{acmart}
\else
	\documentclass[conference]{IEEEtran}
\fi

\usepackage{amsmath}
\DeclareMathOperator*{\argmax}{arg\,max}

\usepackage{multirow}
\usepackage{tabularx,booktabs}
\usepackage{xspace}
\usepackage{pdfcomment}

\usepackage[binary-units=true]{siunitx}
\usepackage{tikz}
\usetikzlibrary{positioning}

\ifacm
	\usepackage[nolist]{acronym}
\usepackage{booktabs} 
\usepackage{subfig}
\usepackage{balance}

\usepackage{array}
\usepackage{graphicx}
\usepackage{wasysym}
\usepackage{xparse}

\usepackage{makecell}
\newcolumntype{Y}{>{\centering\arraybackslash}X}
\hypersetup{draft}

\renewcommand\footnotetextcopyrightpermission[1]{} 
\else
	\usepackage{graphicx}
\usepackage{epstopdf}
\usepackage{standalone}
\usepackage[nolist ]{acronym}
\usepackage{tcolorbox}
\usepackage{pdfpages}
\usepackage[bookmarks=false]{hyperref}
\usepackage{pdfcomment}
\usepackage{hologo}

\usepackage{xstring}

\usepackage[utf8]{inputenc}
\usepackage{marvosym}

\usepackage{algorithm}
\usepackage{algpseudocode}

\usepackage{psfrag}
\usepackage{graphicx}

\usepackage{rotating}

\renewcommand{\baselinestretch}{1}

\usepackage[caption=false,font=footnotesize]{subfig}
\usepackage{stfloats}
\fi

\begin{document}

\newcommand{\paperTitle}{Acting Selfish for the Good of All: Contextual Bandits for Resource-Efficient Transmission of Vehicular Sensor Data}
\newcommand{\paperAuthors}{Benjamin Sliwa and Christian Wietfeld}
\newcommand{\paperEmails}{$\{$Benjamin.Sliwa, Christian.Wietfeld$\}$@tu-dortmund.de}

\newcommand\single{1\textwidth}
\newcommand\double{.48\textwidth}
\newcommand\triple{.32\textwidth}
\newcommand\quarter{.24\textwidth}
\newcommand\singleC{1\columnwidth}
\newcommand\doubleC{.475\columnwidth}

\newcommand{\figurePadding}{0pt}
\newcommand{\figureTopPadding}{\figurePadding}
\newcommand{\figureBottomPadding}{\figurePadding}

\newcommand{\lte}{\ac{LTE}\xspace}
\newcommand{\lena}{\ac{LENA}\xspace}
\newcommand{\ltesim}{LTE-Sim\xspace}
\newcommand{\opnet}{Riverbed\xspace}
\newcommand{\simulte}{SimuLTE\xspace}
\newcommand{\omnet}{\ac{OMNeT++}\xspace}
\newcommand{\inet}{INET\xspace}
\newcommand{\itu}{ITU-R M.2135\xspace}

\newcommand\mno[1]{\emph{\ac{MNO}~#1}\xspace}
\newcommand\red[1]{\colorbox{red}{#1}}
\newcommand\proposal{\ac{BS-CB}\xspace}

\newcommand\tikzFig[2]
{
	\begin{tikzpicture}
		\node[draw,minimum height=#2,minimum width=\columnwidth,text width=\columnwidth,pos=0.5]{\LARGE #1};
	\end{tikzpicture}
}

\newcommand{\dummy}[3]
{
	\begin{figure}[b!]  
		\begin{tikzpicture}
		\node[draw,minimum height=6cm,minimum width=\columnwidth,text width=\columnwidth,pos=0.5]{\LARGE #1};
		\end{tikzpicture}
		\caption{#2}
		\label{#3}
	\end{figure}
}

\newcommand\pos{h!tb}

\newcommand{\basicFig}[7]
{
	\begin{figure}[#1]  	
		\vspace{#6}
		\centering		  
		\includegraphics[width=#7\columnwidth]{#2}
		\caption{#3}
		\label{#4}
		\vspace{#5}	
	\end{figure}
}
\newcommand{\fig}[4]{\basicFig{#1}{#2}{#3}{#4}{0cm}{0cm}{1}}

\newcommand\sFig[2]{\begin{subfigure}{#2}\includegraphics[width=\textwidth]{#1}\caption{}\end{subfigure}}
\newcommand\vs{\vspace{-0.3cm}}
\newcommand\vsF{\vspace{-0.4cm}}

\newcommand{\subfig}[3]
{%
	\subfloat[#3]%
	{%
		\includegraphics[width=#2\textwidth]{#1}%
	}%
	\hfill%
}

\newcommand\circled[1] 
{
	\tikz[baseline=(char.base)]
	{
		\node[shape=circle,draw,inner sep=1pt] (char) {#1};
	}\xspace
}
\begin{acronym}
	\acro{MCS}{Modulation and Coding Scheme}
	\acro{HARQ}{Hybrid Automatic Repeat Request}
	\acro{MAC}{Medium Access Control}
	\acro{TTI}{Transmission Time Interval}
	\acro{TBS}{Transport Block Size}
	\acro{PRB}{Physical Resource Block}
	\acro{CBR}{Constant Bitrate}
	\acro{mMTC}{massive Machine-type Communication}
	\acro{LTE}{Long Term Evolution}
	\acro{UE}{User Equipment}
	\acro{eNB}{evolved Node B}
	\acro{WEKA}{Waikato Environment for Knowledge Analysis}
	\acro{LIMITS}{Lightweight Machine Learning for IoT Systems}
	\acro{RMSE}{Root Mean Squared Error}
	\acro{RF}{Random Forest}
	\acro{MAE}{Mean Absolute Error}
	\acro{DDNS}{Data-driven Network Simulation}
	\acro{MNO}{Mobile Network Operator}
	\acro{AoI}{Age of Information}
	\acro{MTC}{Machine-type Communication}
	\acro{CAT}{Channel-aware Transmission}
	\acro{ML-CAT}{Machine Learning CAT}
	\acro{RL-CAT}{Reinforcement Learning CAT}
	\acro{5GAA}{5G Automotive Association}
	\acro{ITU}{International Telecommunication Union}
	\acro{VSN}{Vehicular Sensor Network}
	\acro{HD}{High Definition}
	\acro{NWDAF}{Network Data Analytics Function}
	\acro{ANN}{Artificial Neural Network}
	\acro{TCP}{Transmission Control Protocol}
	\acro{ECDF}{Empirical Cumulative Distribution Function}
	\acro{SS}{Signal Strength}
	\acro{ASU}{Arbitrary Strength Unit}
	\acro{RSSI}{Reference Signal Strength Indicator}
	\acro{RSRP}{Reference Signal Received Power}
	\acro{RSRQ}{Reference Signal Received Quality}
	\acro{SINR}{Signal-to-interference-plus-noise Ratio}
	\acro{CQI}{Channel Quality Indicator}
	\acro{TA}{Timing Advance}	
	\acro{QoS}{Quality of Service}
	\acro{SDR}{Software Defined Radio}
	\acro{LinUCB}{Linear Upper Confidence Bound}
	\acro{UCB}{Upper Confidence Bound}
	\acro{CART}{Classification and Regression Tree}
	\acro{BS-CB}{Black Spot-aware Contextual Bandit}
	\acro{GPR}{Gaussian Process Regression}
	
\end{acronym}

\title{\paperTitle}

\ifacm
	\newcommand{\cni}{\affiliation{%
		\institution{Communication Networks Institute}
		\city{TU Dortmund University}
		\state{Germany}
		\postcode{44227}\
	}}
	
	\ifdoubleblind
		\author{Anonymous Authors}
		\affiliation{\institution{Anonymous Institutions}}
		\email{Anonymous Emails}

	\else
		\author{Benjamin Sliwa}
		\orcid{0000-0003-1133-8261}
		\cni
		\email{benjamin.sliwa@tu-dortmund.de}
		
		\author{Rick Adam}
		\cni
		\email{rick.adam@tu-dortmund.de}
		
		\author{Christian Wietfeld}
		\cni
	\email{christian.wietfeld@tu-dortmund.de}
	
	\fi

\else

	\title{\paperTitle}

	\ifdoubleblind
	\author{\IEEEauthorblockN{\textbf{Anonymous Authors}}
		\IEEEauthorblockA{Anonymous Institutions\\
			e-mail: Anonymous Emails}}
	\else
	\author{\IEEEauthorblockN{\textbf{\paperAuthors}}
		\IEEEauthorblockA{Communication Networks Institute,	TU Dortmund University, 44227 Dortmund, Germany\\
			e-mail: \paperEmails}}
	\fi
	
	\maketitle

\fi




\begin{abstract}
	
%
%

%
In this work, we present \proposal as a novel client-based method for resource-efficient opportunistic transmission of delay-tolerant vehicular sensor data.
\proposal applies a hybrid approach which brings together all major machine learning disciplines -- supervised, unsupervised, and reinforcement learning -- in order to autonomously schedule vehicular sensor data transmissions with respect to the expected resource efficiency.
%
%
Within a comprehensive real world performance evaluation in the public cellular networks of three \acp{MNO}, it is found that 1) The average uplink data rate is improved by 125\%-195\% 2) The apparently selfish goal of data rate optimization reduces the amount of occupied cell resources by 84\%-89\% 3) The average transmission-related power consumption can be reduced by 53\%-75\% 4) The price to pay is an additional buffering delay due to the opportunistic medium access strategy.

\end{abstract}

\ifacm

	\ccsdesc[300]{Networks~Network resources allocation}
	\ccsdesc[300]{Networks~Network performance modeling}
	\ccsdesc[300]{Networks~Network measurement}
	\ccsdesc[300]{Networks~Mobile networks}
	\ccsdesc[300]{Computing methodologies~Mobile agents}
	\ccsdesc[300]{Computing methodologies~Machine learning}
	\ccsdesc[300]{Computing methodologies~Reinforcement learning}
	\ccsdesc[300]{Computing methodologies~Classification and regression trees}
	
	\keywords{}
\fi

\maketitle
\begin{tikzpicture}[remember picture, overlay]
\node[below=5mm of current page.north, text width=20cm,font=\sffamily\footnotesize,align=center] {Accepted for presentation in: Proceedings of the ACM MobiHoc Workshop on Cooperative Data Dissemination in Future Vehicular Networks (D2VNet)\vspace{0.3cm}\\\pdfcomment[color=yellow,icon=Note]{
@InProceedings\{Sliwa2020,\\
	Author = \{Benjamin Sliwa and Rick Adam and Christian Wietfeld\},\\
	Title = \{Acting Selfish for the Good of All: Contextual Bandits for Resource-Efficient Transmission of Vehicular Sensor Data\},\\
	Booktitle = \{ACM MobiHoc Workshop on Cooperative Data Dissemination in Future Vehicular Networks (D2VNet)\},\\
	Year = \{2020\},\\
\\
	Address = \{Online\},\\
	Month = oct,\\
\}
}};

\end{tikzpicture}

\section{Introduction}

Vehicular crowdsensing \cite{Yu/etal/2018a} is an emerging data acquisition paradigm which utilizes the various sensing and communication capabilities of modern vehicles and exploits their mobility for achieving dynamic sensor coverage of large regions.
%
%
While it is expected that \emph{vehicular big data} will stimulate the development of a multitude of novel data-driven services \cite{Zanella/etal/2014a}, the increase in \ac{mMTC} represents a massive challenge for the cellular network where different users compete among the available cell resources.
%
%
An important observation which motivated our work is the high variance of the resource efficiency of data transmissions along the vehicular trajectories. 
%
%
On the one hand, vehicles encounter periods of high network quality -- also referred to as \emph{connectivity hotspots} -- where data transmissions are performed highly resource efficiently.
%
%
On the other hand, they are also subject to low channel quality periods and encounter network congestion. Here, the mobile \ac{UE} applies a low \ac{MCS} in order to avoid packet errors and retransmissions. Moreover, also the power consumption is often highly increased as the mobile \ac{UE} needs to apply a high transmission power to compensate challenging path loss situations.
Since conventional data transfer methods access the radio medium periodically -- without considering the channel conditions -- a large amount of resources is spend on improving the reliability of the data transfer.
%
%
%

\begin{figure}[]  	
	\vspace{0cm}
	\centering		  
	\includegraphics[width=1.0\columnwidth]{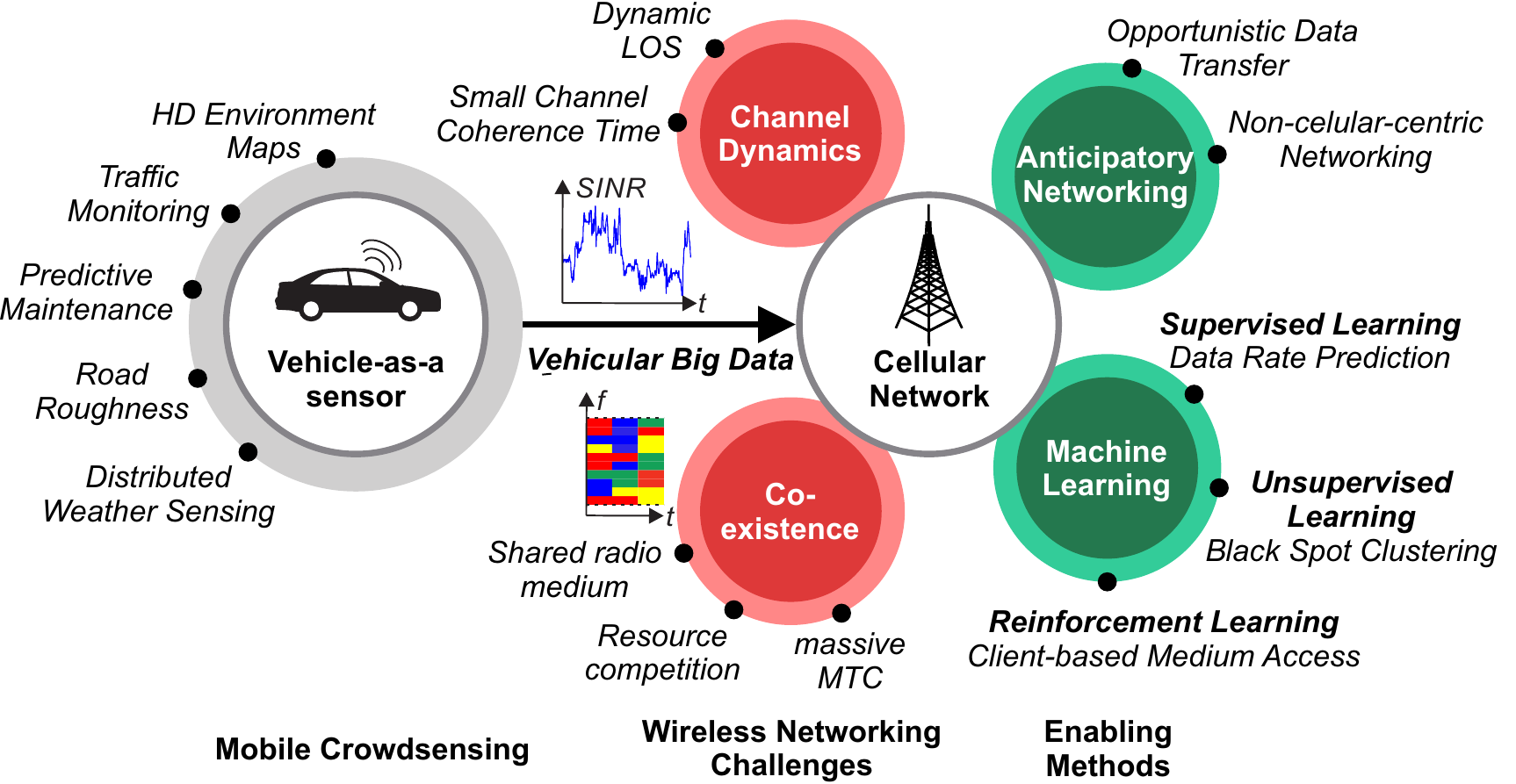}
	\caption{Overview about Applications, Challenges, and Solution Approaches For Vehicular Crowdsensing}
	\label{fig:scenario}
	\vspace{-0.5cm}	
\end{figure}

%
%
\emph{Non-cellular-centric} networking is an emerging research field where client devices become part of the network fabric and participate explicitly or implicitly in network management functions \cite{Coll-Perales/etal/2019a}.
%
%
Client-based \emph{opportunistic} data transfer for \emph{delay-tolerant} applications schedule vehicular sensor data transmissions with respect to the expected resource efficiency: Acquired data is buffered locally until the mobility-dependent channel quality is considered sufficient.
Due to the buffering-related delaying of the data transfer, this approach cannot be applied for safety-criticial data such as cooperate awareness messaging. However, since many vehicle-as-a-sensor applications -- such as updates of \ac{HD} environmental maps and traffic measurements -- allow soft \ac{AoI} deadlines, opportunistic medium access is a promising candidate for utilizing the existing network resources in a more efficient way.
Fig.~\ref{fig:scenario} summarizes the applications, challenges, and solution approaches for vehicular crowdsensing in cellular networks.

%
%

%
%
In this work, we present a novel client-based opportunistic data transmission scheme that relies on a combination of multiple learning models. The contributions are summarized as follows:
%
%
\begin{itemize}
	\item \proposal is a novel \textbf{hybrid machine learning}-enabled transmission scheme for resource efficient transfer of vehicular sensor data.
	\item \textbf{Black spot-aware networking}: Exploitation of knowledge about the geospatially-dependent uncertainties of the prediction model.	
	\item \textbf{Real world} performance evaluation and comparison of the novel approach to existing methods
\end{itemize}

%
%
The remainder of the paper is structured as follows. After discussing the related work in Sec.~\ref{sec:related_work}, we present the proposed \proposal in Sec.~\ref{sec:approach}. Afterwards, an overview about the methodological aspects is given in Sec.~\ref{sec:methods}. Finally, detailed results of real world experiments and data-driven simulations are provided in Sec.~\ref{sec:results}.

\section{Related Work} \label{sec:related_work}

%
%
\textbf{Anticipatory networking} \cite{Bui/etal/2017a} is a novel communications paradigm which aims to optimize decision processes within mobile communication systems through proactive consideration of \emph{context} information.
%
%
Due to the inherent interdependency of mobility and radio propagation dynamics, highly mobile systems such as vehicular networks are expected to benefit significantly from this form of network optimization.
%
%
As pointed out by a recent report of the \ac{5GAA} \cite{5GAA/2020a}, \emph{predictive \ac{QoS}} along the vehicular trajectories will a key enabler for future connected and automated driving.

%
%
\textbf{Machine learning} allows to expose hidden interdependencies between measurable variables and represents a key enabler for anticipatory networking. Machine learning models can be characterized into three major categories:
%
%
\emph{Supervised learning} techniques train a model $f$ on a training data set $\mathbf{X}$ with \emph{labeled} data $\mathbf{Y}$ such that $f: \mathbf{X} \rightarrow \mathbf{Y}$. Afterwards, the trained model can be utilized to make \emph{predictions} on unlabeled data sets.
%
%
\emph{Unsupervised learning} is applied to detect patterns in unlabeled data sets. This allows to cluster data points with similar characteristics, e.g., through application of the popular k-means \cite{Arthur/Vassilvitskii/2007a} method.
%
%
\emph{Reinforcement learning} is an important step towards \emph{zero touch optimization} of wireless communication systems. Hereby, \emph{agents} learn autonomous decision making by performing \emph{actions} within an \emph{environment} through observation of the resulting \emph{rewards}.

A detailed summary about models and applications related to research questions in the wireless communication domain is given by the authors of \cite{Wang/etal/2020a}.
%
%
Within the emerging 5G networks, the integration of machine learning methods mainly focuses on the network infrastructure side. Manifestations of this development can be seen in the \ac{NWDAF} \cite{3GPP/2019a} for network load assessment (e.g., for dynamic slicing) and in the architectural framework defined by the \ac{ITU} \cite{ITU-T/2019a} for utilizing machine learning-based network management.
%
%
It is expected that the trend of replacing mathematical models by machine learning functions will continue further and ultimately lead to \emph{pervasive} machine learning in future networks such as 6G \cite{Ali/etal/2020a}.

%
%
Different research works (e.g., \cite{Herrera-Garcia/etal/2019a, Sliwa/Wietfeld/2019a}) have analyzed 
client-based data rate prediction for mobile networks based on network indicator measurements.
An important observation is that \ac{CART}-based methods such as \acp{RF} \cite{Breiman/2001a} often achieve a better prediction accuracy than more complex methods such as \emph{deep learning} which require a significantly higher amount of training data in order to overcome the \emph{curse of dimensionality} \cite{Zappone/etal/2019a}.

%
%
The advancements in machine learning-enabled networking have also catalyzed the emergence of novel performance analysis methods that focus on end-to-end modeling of wireless communication systems. In this work, we apply a corresponding setup for training and parameterizing the reinforcement learning-based transmission scheme (see Sec.~\ref{sec:methods}):
\ac{DDNS} \cite{Sliwa/Wietfeld/2019c} is a novel machine learning-enabled simulation method which provides fast and accurate modeling of end-to-end performance indicators in concrete evaluation scenarios by replaying empirical context traces. Hereby, multiple prediction models are applied jointly in order to learn the end-to-end behavior of a target performance indicator as well as the statistical derivations between prediction model and ground truth measurements.

\section{Proposed Hybrid Machine Learning Approach} \label{sec:approach}

%
%
\begin{figure}[]  	
	\vspace{0cm}
	\centering		  
	\includegraphics[width=1.0\columnwidth]{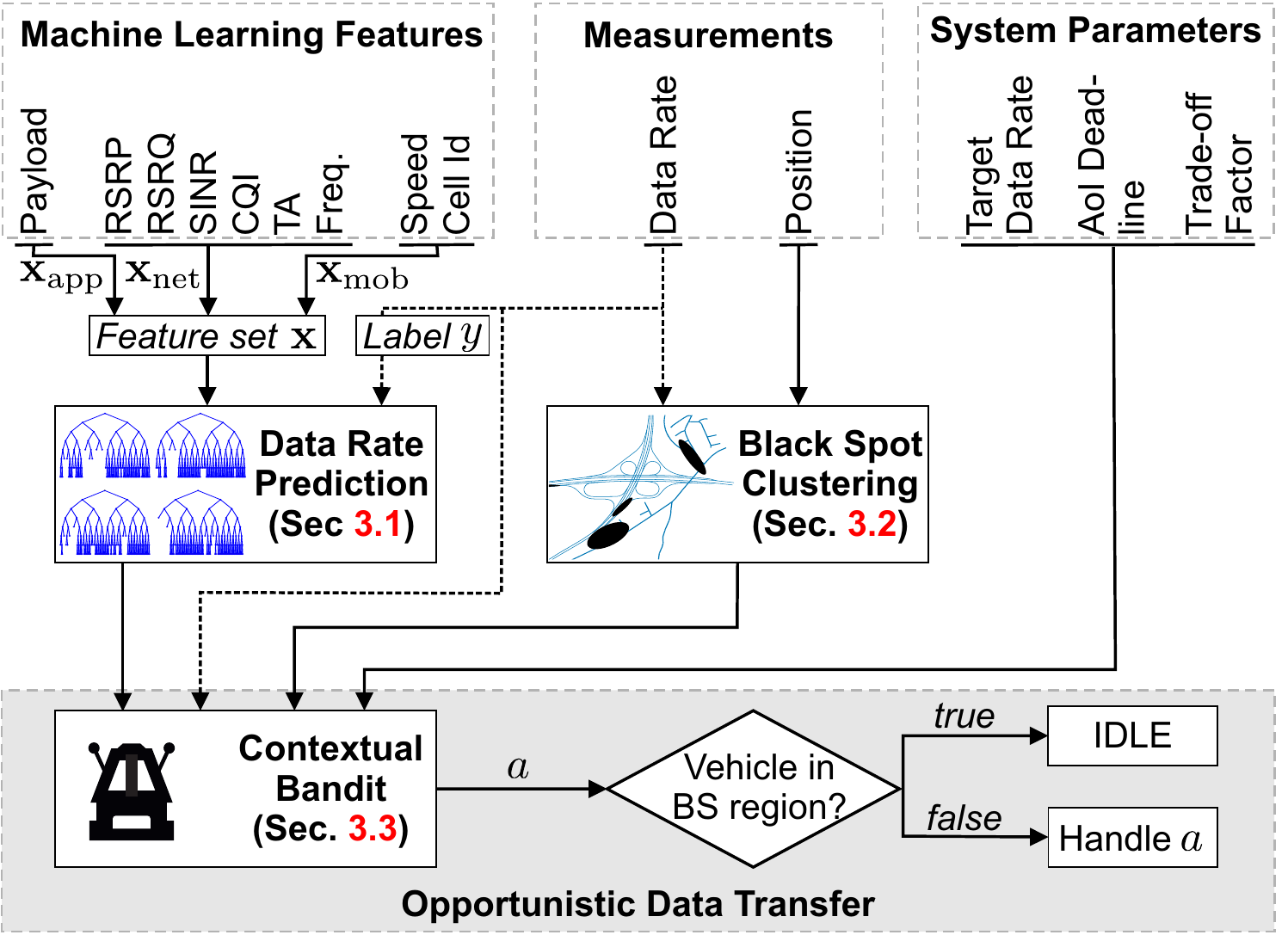}
	\caption{Overall System Architecture Model}
	\label{fig:architecture}
	\vspace{-0.3cm}	
\end{figure}
The overall system architecture model of the proposed solution approach is shown in Fig.~\ref{fig:architecture}.
%
%
Instead of using a multi-dimensional feature vector of raw context measurements for the autonomous decision making, we use an intermediate supervised learning step to forecast the currently achievable data rate in order to reduce the dimensionality of the learning problem.
Moreover, knowledge about the geospatial dependency of the prediction errors is utilized to improve the opportunistic data transfer process.
In the following, the different modules are explained in further details.

\subsection{Supervised Learning for Data Rate Prediction}

The overall feature set $\mathbf{x}$ is composed of measurements from different context domains
\begin{itemize}
	\item \textbf{Network features} $\mathbf{x}_{\text{net}}$: \ac{RSRP}, \ac{RSRQ}, \ac{SINR}, \ac{CQI}, \ac{TA} and carrier frequency
	\item \textbf{Mobility features} $\mathbf{x}_{\text{mob}}$: Speed of the vehicle and cell id of the connected \ac{eNB}
	\item \textbf{Application features }$\mathbf{x}_{\text{app}}$: Payload size of the data packet to be transmitted
\end{itemize}
Due to the findings of the in-depth comparison of different data rate prediction models in \cite{Sliwa/Wietfeld/2019c}, we apply a \ac{RF} model for predicting the currently achievable data rate as $\tilde{S} = f_{\text{RF}}(\mathbf{x})$.

\subsection{Unsupervised Learning for Black Spot Clustering}

In previous work \cite{Sliwa/Wietfeld/2020a}, we have pointed out that the achievable accuracy of prediction models has a \emph{geospatial dependency}: Artifacts in the observed prediction performance often occur cluster-wise and are mostly related to effects which are not covered by the feature set (e.g., handovers, short term link loss).
Although this knowledge does not allow us to compensate the undesired effects, it can be utilized as a measurement of \emph{trust} into the prediction model in order to strengthen the robustness of the context-aware data transfer.
%
%
With respect to its usage in traffic safety, where the term \emph{black spot} corresponds to a geographical region with an increased probability for collisions, we migrate its usage to the wireless communications domain and use it as a description for geographical regions with exceptional high prediction uncertainty.
%
%
%

\begin{figure}[] 
	\centering
	\subfig{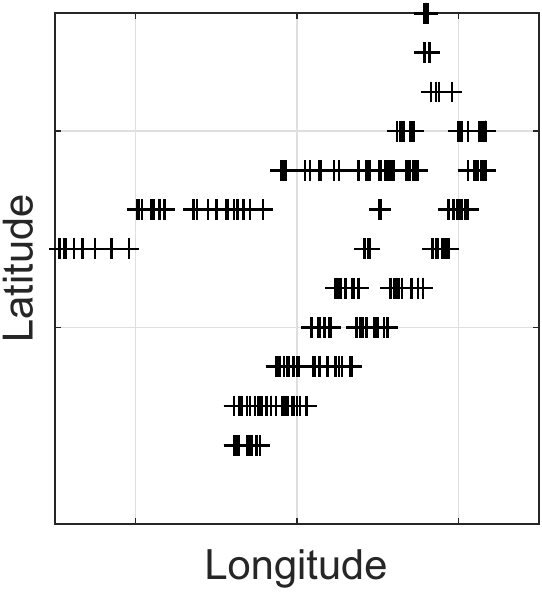}{0.15}{Raw Measurements}%
	\subfig{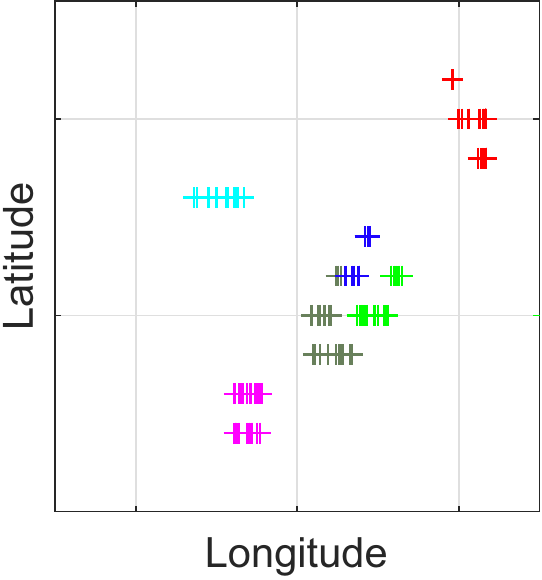}{0.15}{Clustered Black Spot Measurements}%
	\subfig{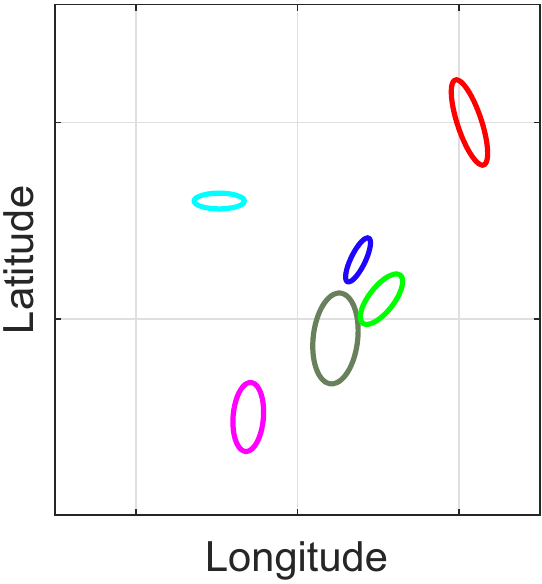}{0.15}{Fitted Ellipses}%
	\caption{Steps for the Determination of Black Spot Regions}
	\label{fig:clustering}
\end{figure}

%
%
The black spot-aware approach is divided into two phases:

%
%
\textbf{Offline data analysis:}
%
%
At first, \emph{k-means} \cite{Arthur/Vassilvitskii/2007a} is applied to perform a geo-spatial clustering of the data points into a total amount of $N_{c}$ clusters.
For each cluster $c$ with $N$ cluster points, the \ac{RMSE} is calculated based on the difference between predictions $\tilde{{S}}$ and measurements ${S}$ as
\begin{equation}
	\text{RMSE} = \sqrt{\frac{\sum_{i=1}^{N} \left(  \tilde{S}_{i} - S_{i} \right)^2}{N}}.
\end{equation}
%
%
If the computed value exceeds a defined threshold $\text{RMSE}_{\max}$, the cluster $c$ is considered as a  \emph{black spot cluster}.
%
%
Finally, all black spots clusters are fitted to ellipses based on the dominant intra-cluster distance vector.
Fig.~\ref{fig:clustering} summarizes different steps for of the black spot cluster determination.

\begin{figure}[]  	
	\vspace{0cm}
	\centering		  
	\includegraphics[width=0.9\columnwidth]{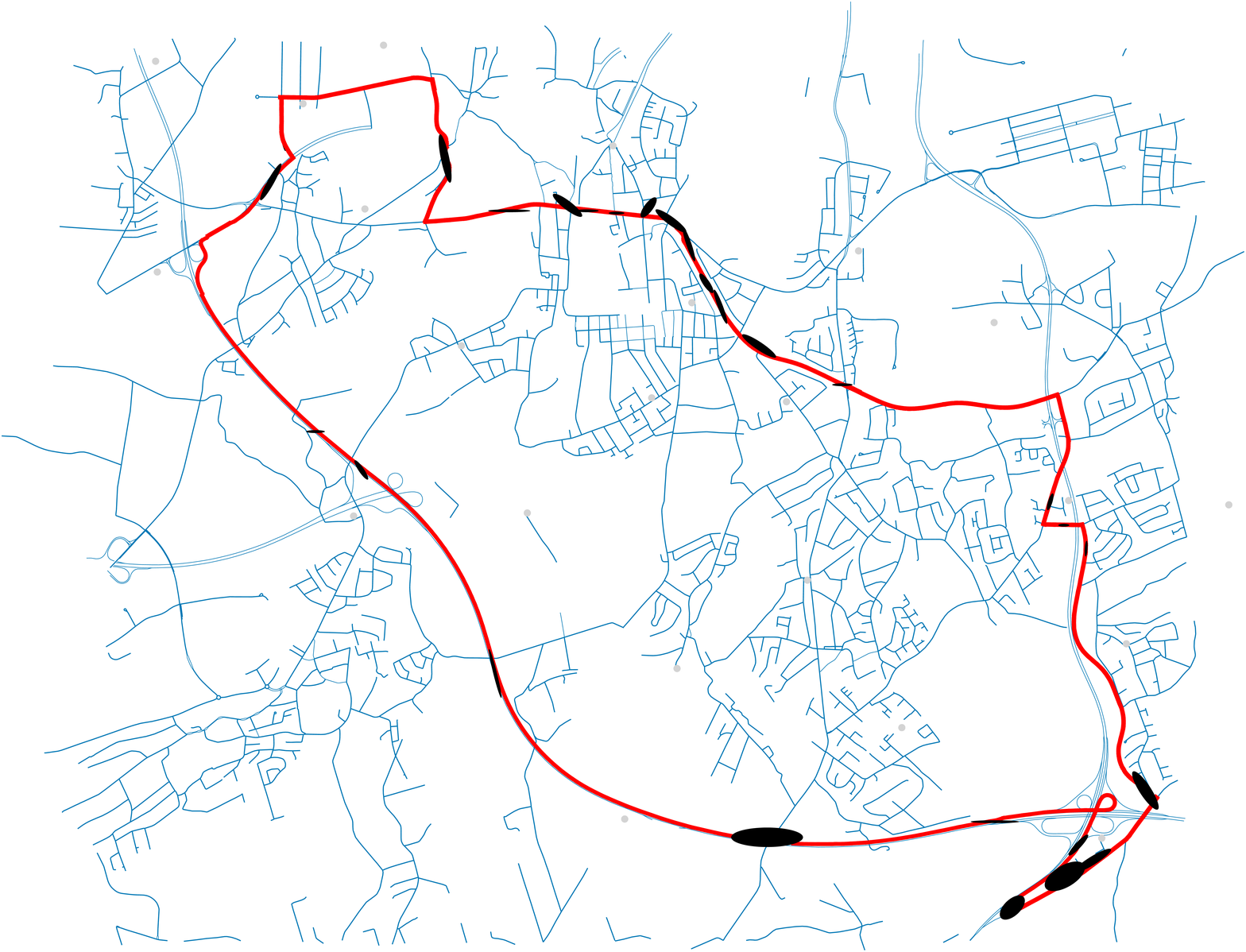}
	\caption{Resulting Black Spot Regions for \mno{A} on the Evaluation Track (Map: ©OpenStreetMap contributors, CC BY-SA)}
	\label{fig:map}
	\vspace{-0.6cm}	
\end{figure}

%
%
\textbf{Online application}: For the later exploitation of the derived knowledge by the reinforcement learning-based data transmission, a vehicle needs to know if it is currently within a black spot region.
%
%
For a given cartesian point $\mathbf{P}$, an intersection test for an $\alpha$-rotated ellipse centered at $\mathbf{P_0}$ is performed as
%
%
\begin{equation}
	\frac{(c \cdot \mathbf{v}.x + s \cdot \mathbf{v}.y)^2}{a^2}
	+ \frac{(s \cdot \mathbf{v}.x - c \cdot \mathbf{v}.y)^2}{b^2}
	\leq 1
\end{equation}
with $\mathbf{v} = \mathbf{P} - \mathbf{P_0}$, $c = \cos \alpha$, and $s = \sin \alpha$.
An example for the black spot regions for \mno{A} on the considered evaluation track is shown in Fig.~\ref{fig:map}.

\subsection{Reinforcement Learning for Opportunistic Data Transfer} \label{sec:opportunistic_data_transfer}

%
%
The actual opportunistic data transfer process is represented by a \ac{LinUCB} \cite{Li/etal/2010a} contextual bandit with two \emph{arms} which correspond to the possible \emph{actions}:
%
%
\begin{itemize}
	\item $\mathbf{a}_{\textbf{IDLE}}$ delays the data transfer in favor of an expected resource efficiency improvement in the future. Acquired sensor data is buffered locally.
	\item $\mathbf{a}_{\textbf{TX}}$ transmits the whole data buffer.
\end{itemize}

%
%
The context-aware arm selection process is modeled as 
%
%
\begin{equation}
	a_t = \argmax_{a\in \mathbf{A}_{t}} \left(
	\underbrace{\hat{\theta}^{T}_{a} \mathbf{x}_{t,a}}_{\text{Estimated reward}} + 
	\underbrace{\alpha \sqrt{\mathbf{x}^{T}_{t,a} \mathbf{A}^{-1}_{a} \mathbf{x}_{t,a}}}_{\text{UCB}~\mathbf{C}_{a}}
	\right) 
\end{equation}
%
%
whereas the estimated arm reward is derived through ridge regression with $\hat{\theta}_{a}$ being the regression coefficients and $\mathbf{x}_{t,a} = \{\tilde{S}(t), \Delta t\}$ being the $d$-dimensional feature vector for arm $a$ in time step $t$. The parameter $\alpha = 1 + \sqrt{\frac{\ln(2/\delta)}{2}}$ controls the degree of exploration based on the only system parameter $\delta$.
%
%
For the \ac{UCB} part, $\mathbf{A}_{a} = \mathbf{D}^{T}_{a}\mathbf{D}_{a}+\mathbf{I}_{a}$ consists of a $d$-dimensional identity matrix $\mathbf{I}_{a}$ and $\mathbf{D}_{a}$ as a $m \times d$ matrix that contains the $m$ rows of training inputs.

%
%
%
%
After performing either the \texttt{TX} or the \texttt{IDLE} action, a real-valued reward $r_{t}$ is observed and the regression coefficients are updated as:
%
%
\begin{equation}
\hat{\theta}_{a} \leftarrow \mathbf{A}^{-1}_{a} \mathbf{b}_{a}
\end{equation}
with 
%
%
\begin{equation}
	\mathbf{b}_{a_{t}} \leftarrow \mathbf{b}_{a_{t}} + r_{t} \mathbf{x}_{t,a_{t}}
\end{equation}
whereas $\mathbf{b}_{a_{t}}$ is set to a $d$-dimensional zero vector upon first initialization. The reward is calculated action-specific based on the corresponding reward functions:
%
%
\begin{equation}
		r_{\text{TX}}(S, \Delta t) = \frac{\omega \cdot (\tilde{S}-S^{*})}{S_{\max}} + \frac{\Delta t \cdot (1-\omega)}{\Delta t_{\max}} 
\end{equation}
%
%
%
%
%
\begin{equation}
	r_{\text{IDLE}}(\Delta t) = \begin{cases}
	\Omega & \Delta t \geq \Delta t_{\max} \\ 
	0 & \text{else}
	\end{cases}
\end{equation}
whereas $S^{*}$ represents an \ac{MNO}-specific target data rate and $\Delta t_{\max}$ corresponds to an application-specific upper bound for the tolerable \ac{AoI}. $w$ is a trade-off parameter for controlling the focus on either data rate optimization or \ac{AoI} focus. $\Omega$ is a negative number which is used as a \emph{deadline violation punishment} in order to ensure that the \texttt{TX} action is immediately if the deadline is violated.

\section{Methodology} \label{sec:methods}

A two-state methodological approach is applied: At first, a \ac{DDNS} setup (see \cite{Sliwa/Wietfeld/2019c}) is utilized to train the reinforcement learning mechanism. Afterwards, we perform a real world measurement study for comparing the novel approach with different existing methods:
\begin{itemize}
	\item \textbf{Periodic transfer} represents the typical \ac{MTC} approach where data is transmitted based on a fixed interval (here $\Delta t=10s$) without considering the current channel quality.
	\item \textbf{\acf{CAT}} \cite{Ide/etal/2015a} is a probabilistic data transmissions scheme which uses the measured \ac{SINR} for client-side scheduling of sensor data transmissions.
	\item \textbf{\acf{ML-CAT}} \cite{Sliwa/etal/2019d} is a machine learning-based extension to \ac{CAT}. Instead of only using a single network quality indicator for the opportunistic medium access, \ac{ML-CAT} uses the predicted data rate (similar to Sec.~\ref{sec:approach})
	\item \textbf{\acf{RL-CAT}} \cite{Sliwa/Wietfeld/2020a} is a first reinforcement learning-enabled data transfer method which replaces the probabilistic medium access with \emph{Q-learning}-based decision making.
\end{itemize}
For the real world evaluation, we consider a 25 km long evaluation track which consists of highway and suburban parts. For each transmission scheme, five drive tests are performed where sensor is transmitted via \ac{TCP} in the uplink through the cellular network of three different German \acp{MNO}. All transmissions are performed with an Android-based \ac{UE} (Samsung Galaxy S5 Neo, Model SM-G903F). The applied \proposal parameters are summarized in Tab.~\ref{tab:parameters}.
%
%
\begin{table}[ht]
	\centering
	\vspace{-0.1cm}
	\caption{Default parameters of the evaluation setup}
	\vspace{-0.2cm}
	\begin{tabular}{ll}
		\toprule
		\textbf{Parameter} & \textbf{Value} \\

		\midrule
	
		Maximum buffering time $\Delta t_{\max}$ & 120~s \\
		Trade-off factor $w$ & 0.9 \\
		Deadline violation punishment $\Omega$ & -1 \\
		Exploration parameter $\delta$ & 0.1 \\
		Number of clusters $N_{c}$ & 100 \\
		MNO-specific black spot threshold $\text{RMSE}_{\max}$ & 3, 2.25, 2.5 \\
		
		\bottomrule
		
	\end{tabular}
	\vspace{-0.3cm}
	\label{tab:parameters}
\end{table}

%
%
The prediction models are learned with the \ac{WEKA}-based \cite{Hall/etal/2009a} \ac{LIMITS} \cite{Sliwa/etal/2020c} framework which provides automatic generation of \texttt{C/C++} code for the trained models.
For unsupervised learning and the \ac{GPR} models required for the \ac{DDNS} setup, the \emph{Statistics and Machine Learning Toolbox} of \texttt{MATLAB} is utilized.

%
%
For analyzing the communication-related power consumption of the \ac{UE}, the most important indicator is the applied transmission power $P_{\text{TX}}$.
Although \texttt{Android}-based \acp{UE} do not expose this information to the user space, it can be inferred from radio signal measurements due to a significant correlation with distance-dependent indicators such as \ac{RSRP} \cite{Falkenberg/etal/2018a}. In order to determine the power consumption as a function of the applied transmission power, we utilize laboratory measurements of the device-specific power consumption behavior. A deeper discussion about the applied method can be found in \cite{Sliwa/etal/2019d}.

%
%
For calculating the network resource efficiency of the transmission schemes in the post processing, we revert the table lookup procedure described in \cite{Satoda/etal/2020a}. Based on the \ac{CQI} measurements, the required \ac{MCS} and \ac{TBS} indices are obtained from a lookup table.

\section{Results} \label{sec:results}

In this section, the results for the \ac{DDNS}-based system optimization as well as for the real world performance evaluation are presented.

%
%
\begin{figure}[]  	
	\vspace{0cm}
	\centering		  
	\includegraphics[width=1.0\columnwidth]{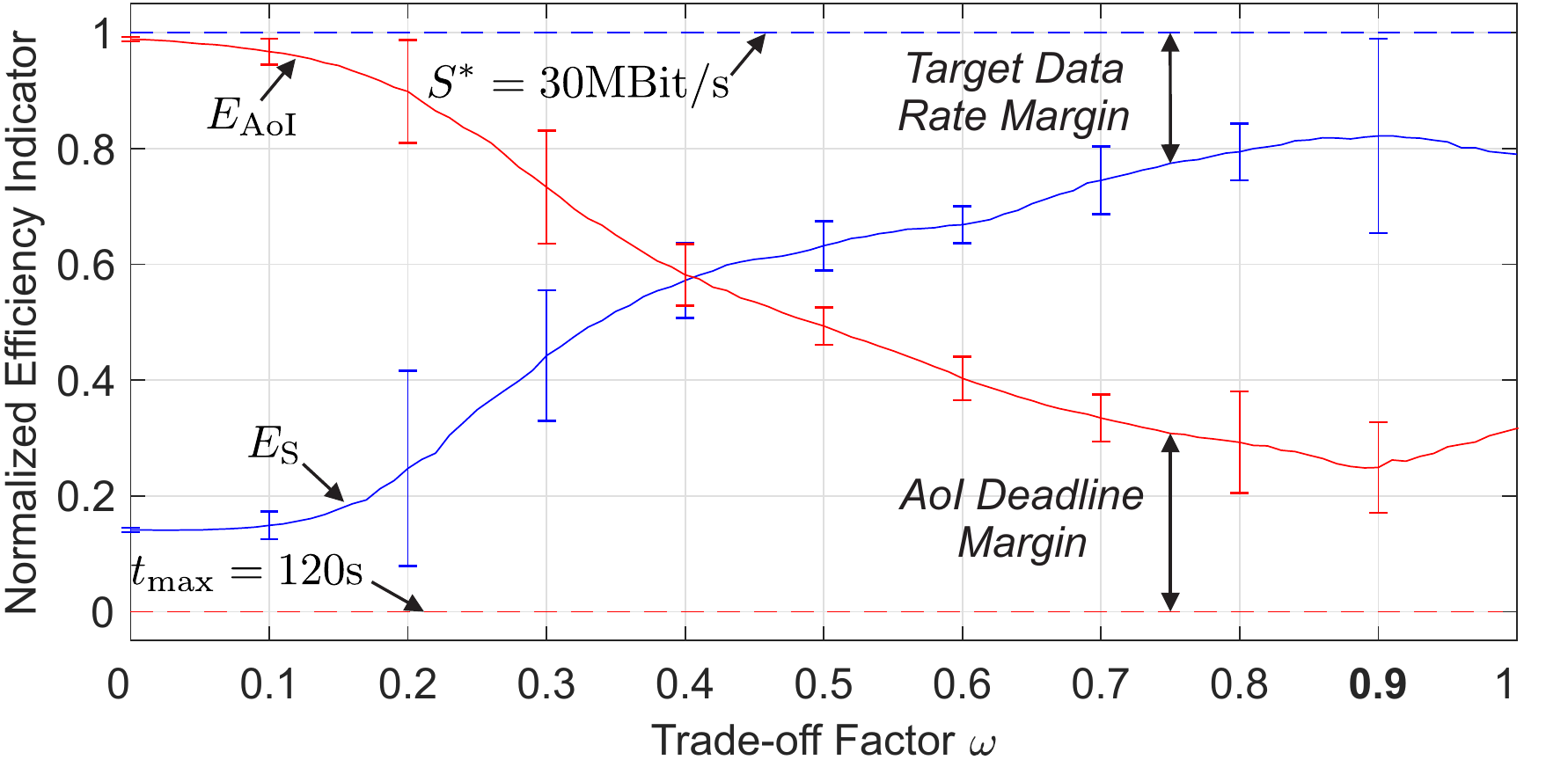}
	\vspace{-0.5cm}	
	\caption{Controllable Trade-off Between Data Rate and \ac{AoI} Optimization}
	\label{fig:tradeoff}
	\vspace{-0.3cm}	
\end{figure}
%
%
%

%
%
\begin{figure}[]  	
	\vspace{0cm}
	\centering		  
	\includegraphics[width=1.0\columnwidth]{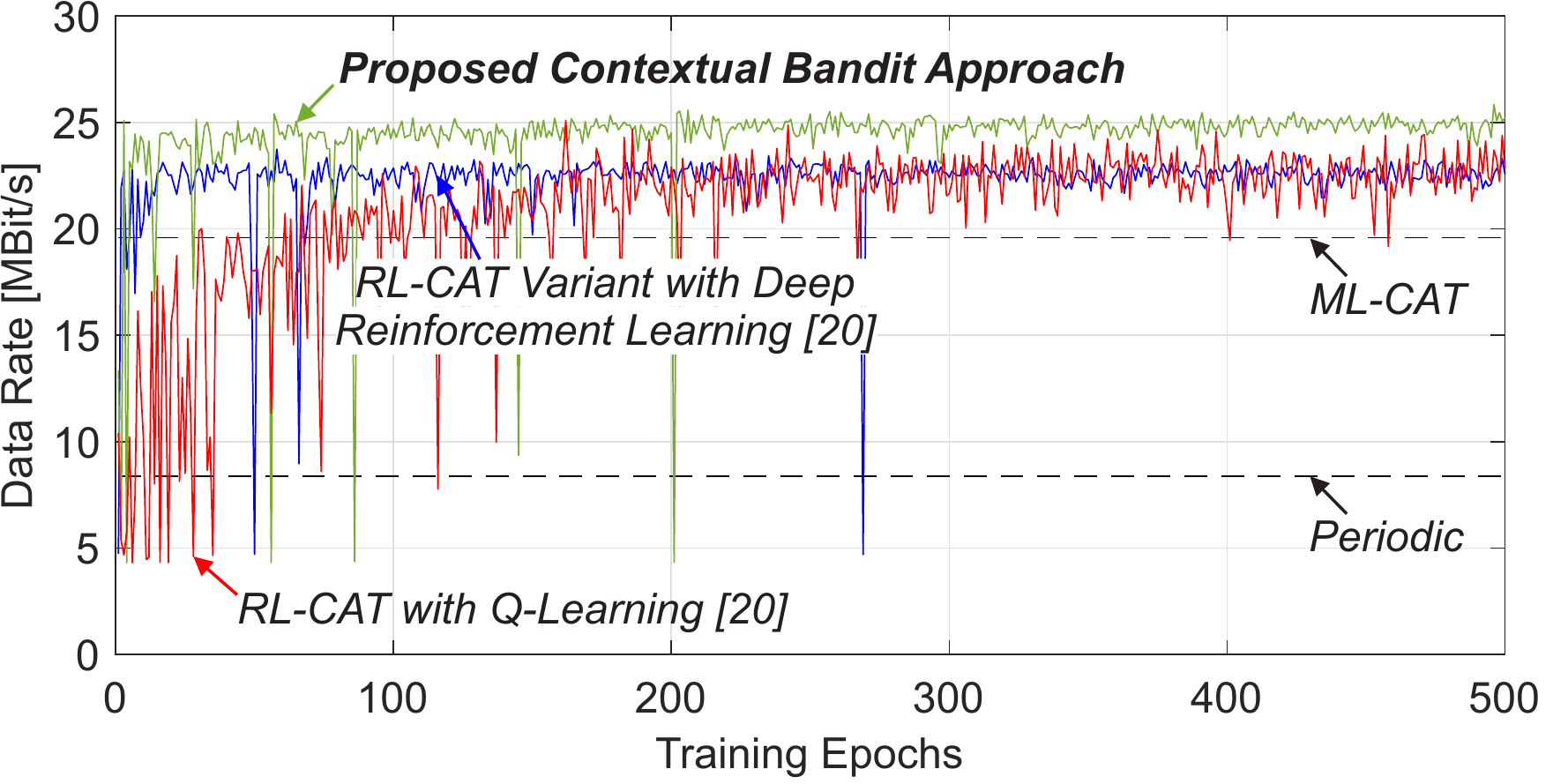}
	\vspace{-0.7cm}	
	\caption{Convergence of the Reinforcement Learning Process}
	\label{fig:convergence}
	\vspace{-0.5cm}	
\end{figure}
%
%
%

%
%
\begin{figure*}[] 
	\centering
	\subfig{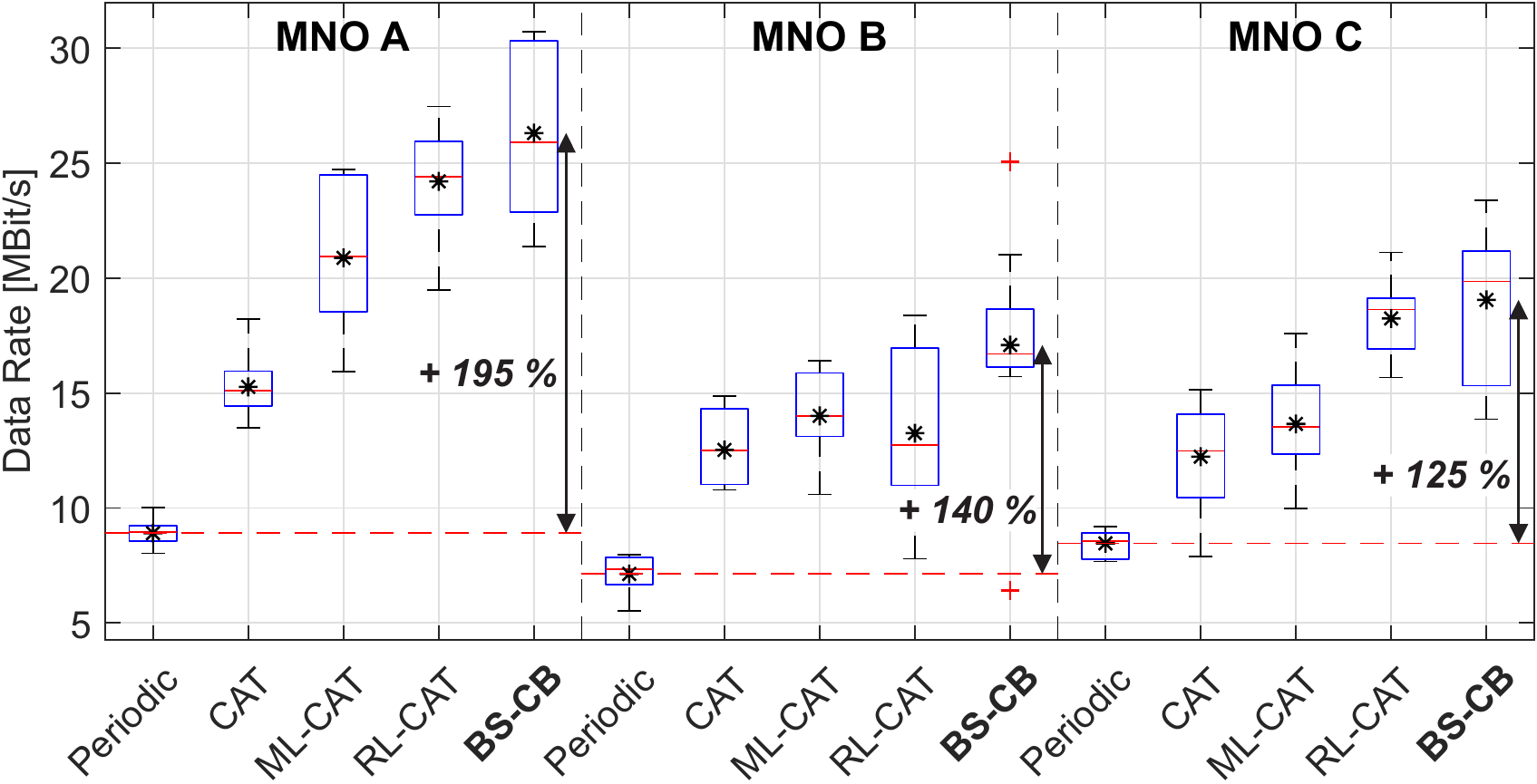}{0.48}{}%
	\subfig{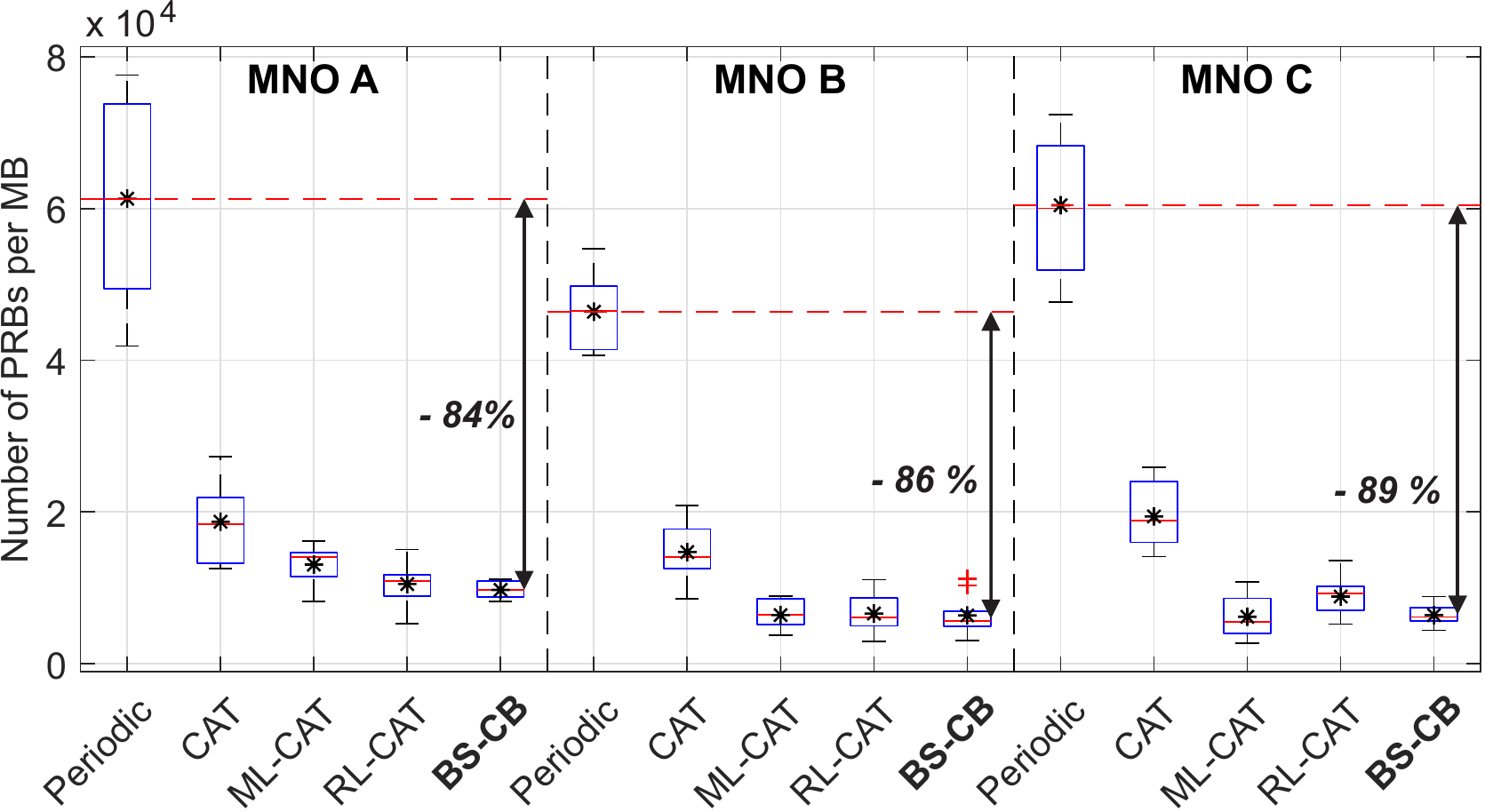}{0.48}{}%
	\vspace{-0.3cm}
	\subfig{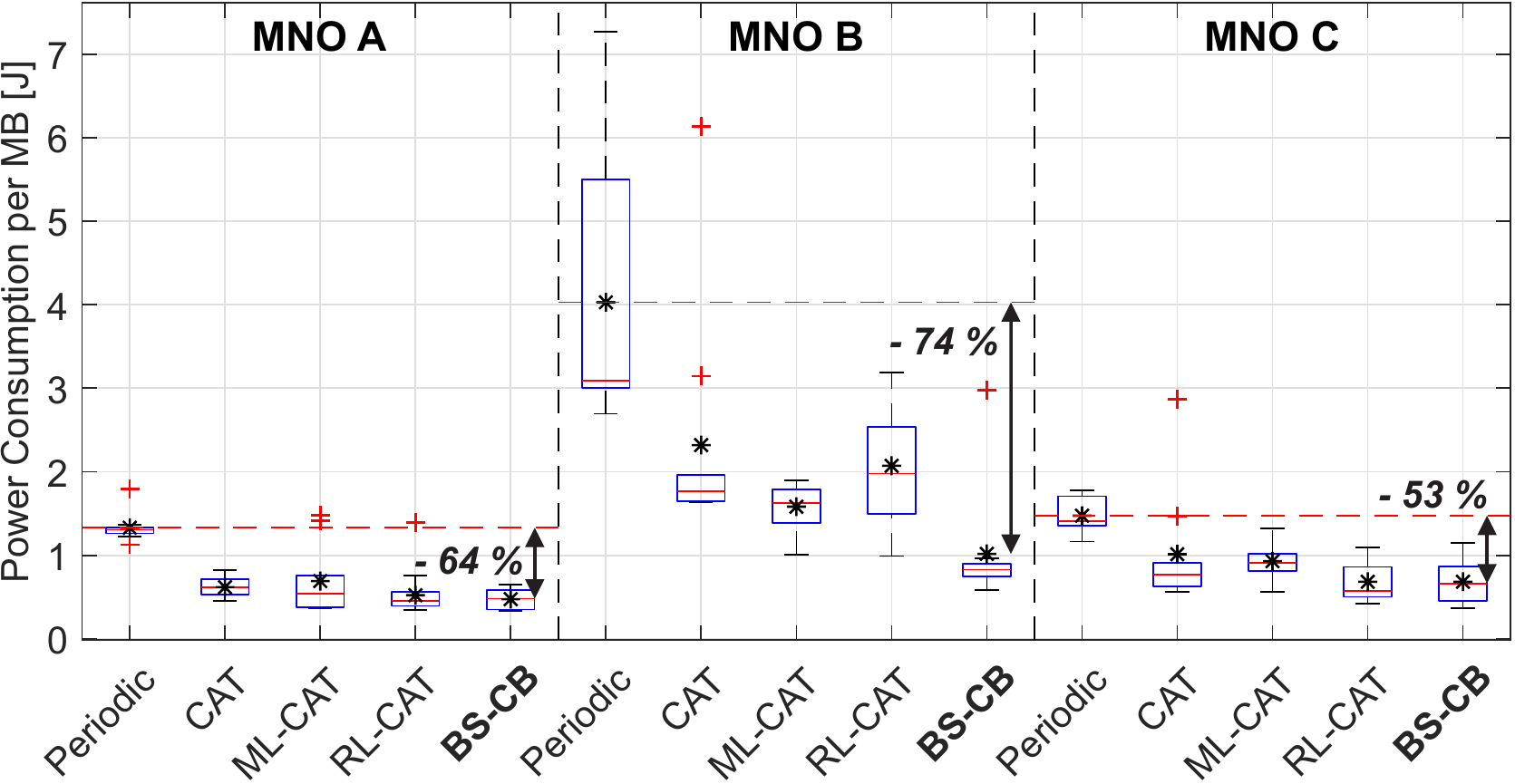}{0.48}{}%
	\subfig{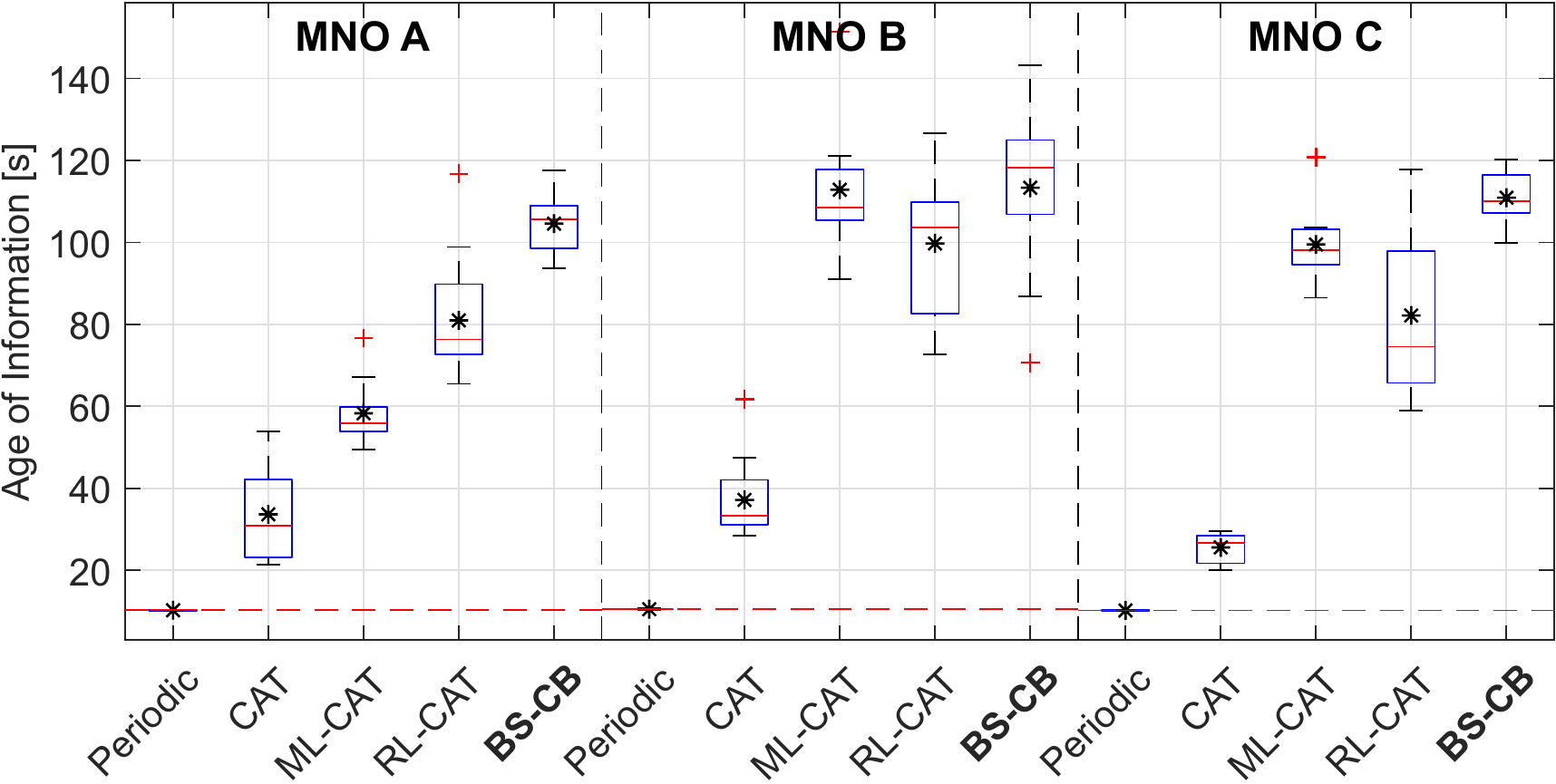}{0.48}{}%
	\vspace{-0.5cm}
	\caption{Performance Comparison of Opportunistic Transmission Schemes for Multiple \acp{MNO}}
	\vspace{-0.3cm}
	\label{fig:data_rate_boxplot}
\end{figure*}

\subsection{Parameterization and Convergence}

%
%
As discussed in Sec.~\ref{sec:opportunistic_data_transfer}, opportunistic data transfer is subject to a fundamental trade-off between data rate and \ac{AoI} optimization which can be controlled via the trade-off factor $w$. For the purpose of comparing the performance in both dimensions, we define two efficiency indicators:
\begin{itemize}
	%
	%
	\item The \textbf{data rate efficiency} $E_{\text{S}} = \bar{S} / S^{*}$ measures how good the average data rate $\bar{S}$ approaches the target data rate $S^*$
	%
	%
	\item The \textbf{\ac{AoI} efficiency} 	$E_{\text{AoI}} = 1- \bar{\Delta t} / \Delta t_{\max}$ is a measure for the margin between the average \ac{AoI} and the deadline $\Delta t_{\max}$ 
\end{itemize}
Fig.~\ref{fig:tradeoff} shows the normalized behavior of both indicators for different values of $w$. 
%
%
Is can be seen that the data rate benefits from larger packets -- which correspond to a lower \ac{AoI} efficiency -- in order to achieve a better payload-overhead ratio and a better compensation of the slow start mechanism of \ac{TCP}.
%
%
%
%
%
In the following, we focus our analysis on data rate optimization and assume $w=0.9$.

%
%
Before the novel transmission scheme can be efficiently applied in the real world, the reinforcement learner needs to adjust its decision making through observation of a multitude of performed transmissions. For this purpose, we replay the measurements of \cite{Sliwa/Wietfeld/2019a} offline. Hereby, each \emph{epoch} represents one virtual drive test on the evaluation track within the \ac{DDNS}.
Fig.~\ref{fig:convergence} shows the resulting data rate of the proposed contextual bandit-based transmission scheme.
%
%
For reference, the convergence behavior of a Q-learning approach according to \cite{Sliwa/Wietfeld/2020a} and a deep reinforcement learning variant of the latter are shown. Hereby, the corresponding \ac{ANN} is set up according to \cite{Sliwa/Wietfeld/2019c} with two hidden layers and ten neurons per hidden layer.
%
%
It can be seen that the proposed contextual bandit-based method achieves the highest absolute data rate and provides an early convergence which is reached after $\sim$200 epochs.
%
%
For the considered deep reinforcement learning and Q-learning methods, the final data rate of the converged system is significantly lower. Moreover, the Q-learning based approach shows a slow convergence behavior.

\subsection{Real World Performance Comparison}

The performance of the converged transmission schemes is now analyzed in a real world scenario (see Sec.~\ref{sec:methods}). Fig.~\ref{fig:data_rate_boxplot} shows multiple performance indicators for the proposed transmission scheme as well as for the considered references.
%
%
It can be observed that the resulting data rate is continuously improved through the different evolution stages of opportunistic data transfer: While the \ac{SINR}-aware \ac{CAT} method already outperforms the periodic approach, the introduction of machine learning-based network quality assessment by \ac{ML-CAT} leads to significant performance improvement. Ultimately, reinforcement learning-based autonomous decision making (\ac{RL-CAT} and \proposal) achieves the highest data rate values. For \mno{A}, \proposal almost triples the resulting data rate.
%
%
In addition, it can be seen that the apparently selfish goal of data rate optimization results in a significant reduction of \ac{MTC}-related resource occupation -- 84\% to 89\% -- which contributes to a better overall coexistence of different resource-consuming entities within the network.
%
%
As a side effect, also the power consumption of the mobile \ac{UE} is reduced as the opportunistic transmission approaches implicitly prefer higher \ac{RSRP} values which have a strong correlation with the applied transmission power \cite{Falkenberg/etal/2018a}. 
%
%
For \mno{B}, it can be seen that the general power consumption level is much higher than for the other \acp{MNO}. In this scenario, the average distance to the \acp{eNB} is significantly higher for \mno{B} then for the other \acp{MNO}. As a result, a significantly higher transmission power is applied, which causes the mobile \ac{UE} to be in a less power-efficient amplification stage for most of the time \cite{Falkenberg/etal/2018a}.
%
%
While the previous results have shown that opportunistic sensor data transfer allows to achieve significant improvements on the client and network side, the price to pay is an increased \ac{AoI} -- about nine times the \ac{AoI} of the periodic approach -- which is the result of the buffering delay. However, the proposed method allows to specify an upper limit for the acceptable \ac{AoI} via the parameter $\Delta t_{\max}$ (see Sec.~\ref{sec:opportunistic_data_transfer}).

\subsection{Side Effects of Black Spot-aware Communication}

Since the black spot-aware data transfer avoids transmissions if the \ac{UE} is within a black spot region, it causes an additional buffering delay. Therefore, we now investigate the times and distances the vehicles spend within the black spot regions.
%
%
\begin{figure}[] 
	\centering
	\subfig{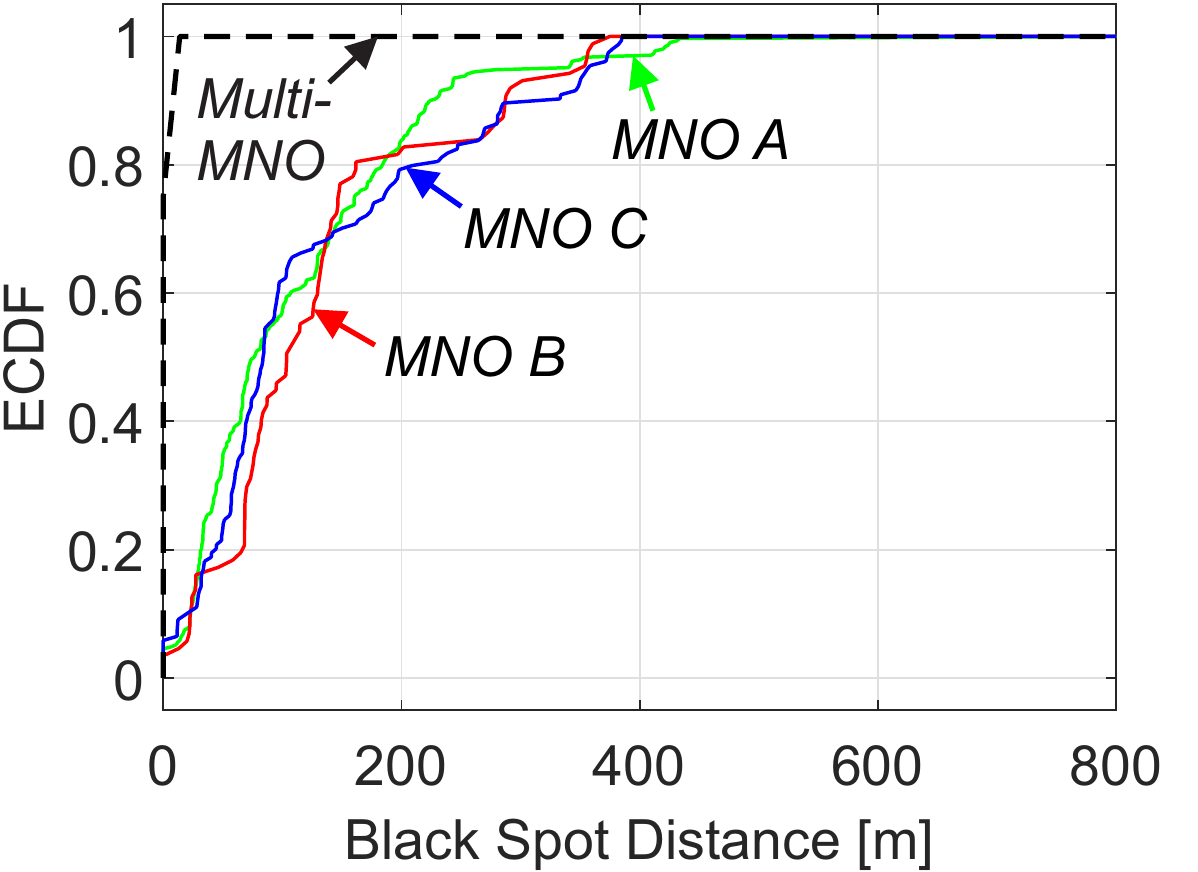}{0.23}{}%
	\subfig{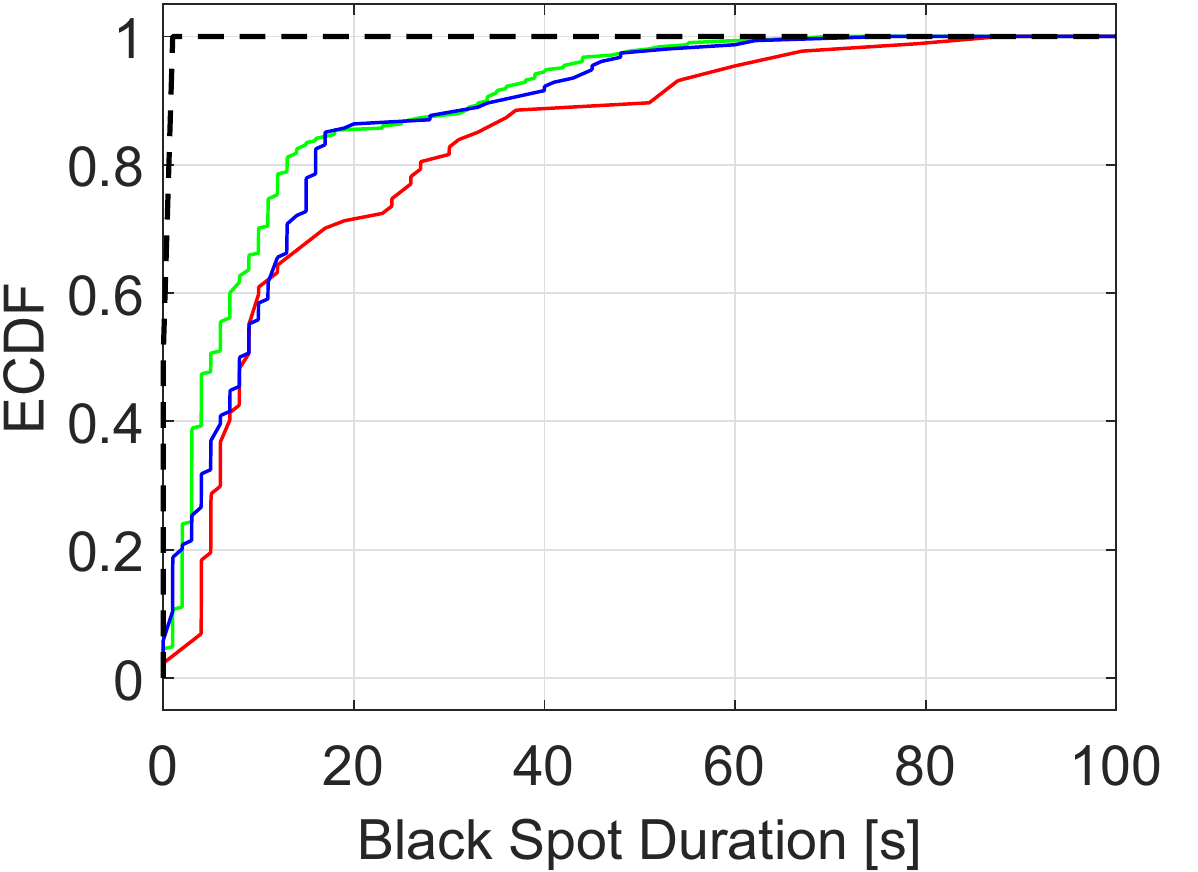}{0.23}{}%
	\vspace{-0.5cm}
	\caption{Black Spot Statistics}
	\vspace{-0.5cm}
	\label{fig:bs_statistics}
\end{figure}
Fig.~\ref{fig:bs_statistics} shows the corresponding \acp{ECDF} for the three \acp{MNO}. In addition, the behavior of a potential future multi-MNO extension are shown where the vehicle dynamically changes the network if it is within a black spot region.
For all \acp{MNO}, 50~\% of the black spot regions spread no more than 100~m which only results in a slight additional delay. 
%
%
However, within the considered scenario, most of the black spots could be compensated through a multi-\ac{MNO} approach which massively reduces the side effects of the black spot-aware approach.
\section{Conclusion}

%
%
In this paper, we presented \proposal as a novel approach for opportunistic data transfer for vehicular sensor data. The proposed method makes use of a hybrid machine learning approach: Reinforcement learning is applied to autonomously schedule data transmissions with respect to the network quality based on data rate predictions. In addition, knowledge about geographically clustered black spot regions is utilized for avoiding transmissions with high prediction uncertainties.
%
%
In a comprehensive real world evaluation, it was shown that the novel method not only achieves significant improvements for the uplink data rate and power consumption of the mobile \ac{UE}, but also contributes to optimizing the resource efficiency of delay-tolerant \ac{MTC} applications.
%
%
In future work, we want to extend \proposal with a multi-\ac{MNO} strategy which allows dynamic network selection  for compensating black spots regions.
%
%
In addition, we plan to further analyze cooperative approaches -- where the network infrastructure actively distributes network load information to the mobile clients \cite{Sliwa/etal/2020b} -- for data rate prediction in order to optimize the resulting accuracy. 
Moreover, we aim to move another step forward towards zero touch optimization through integration of online learning mechanisms for the data rate prediction. This would then allow the system to self-adapt to the \emph{concept drift} caused by significant changes within the cellular network.

\ifdoubleblind

\else

	\section*{Acknowledgment}
	
	\footnotesize
	This work has been supported by the German Research Foundation (DFG) within the Collaborative Research Center SFB 876 ``Providing Information by Resource-Constrained Analysis'', project B4.

\fi

\ifacm
	\bibliographystyle{ACM-Reference-Format}
	\bibliography{Bibliography}
\else
	\bibliographystyle{IEEEtran}
	\bibliography{Bibliography}
\fi

\end{document}